# Fractional motions of an active particle on the quantum vortex

Yun Jeong Kang [1], Sung Kyu Seo [2] and Kyungsik Kim [3,4,*]

[1] School of Liberal Studies, Wonkwang University, Iksan, 54538, Republic of Korea
[2] Haena Ltd., Seogwip0-si, Jeju-do 63568, Republic of Korea
[3] DigiQuay Ltd., Manan-Gu Anyang, Gyeonggido 14084, Republic of Korea
[4] Department of Physics, Pukyong National University, Busan 48513, Republic of Korea

**Abstract**

We analytically investigate the diffusive motion inferred from experimental observations of active particles driven by quantum vortices on the surface of superfluid helium. We first study the dynamical behavior of an active particle subject to a viscoelastic memory effect characterized by a power-law kernel $\mid (t-t')/\tau_{\text{th}} \mid^{-2\beta-1}$, where $\tau_{\text{th}}$ denotes the thermal characteristic time, in the presence of thermal noise. We then analyze the dynamics of an active particle under a uniform vortex force, thermal noise, and viscous dissipation subject to a power-law kernel. Next, by including a harmonic confining force, we obtain analytical solutions for the joint probability density in two distinct time regimes: $t \ll \tau_{\text{th}}$, and $t \gg \tau_{\text{th}}$. As a result, for $\beta = 0.65 \sim 0.7$, the mean-squared displacement exhibits anomalous diffusion with exponent $\alpha = 1.6 \sim 1.7$, defined by $<x_{th}^2(t)> \sim t^\alpha$, in excellent agreement with experimental observations. For $\beta = 1$, the mean-squared displacement reduces to the normal diffusive form with $\alpha = 1.0$. On the other hand, in the presence of a harmonic force and thermal noise, the higher-order moments scale as $t^{4-4\beta}$ in the long-time limit, whereas for particles driven solely by thermal noise they exhibit superdiffusive scaling as $t^{2-4\beta}$ in both the short- and long-time regimes, $t \ll \tau_{\text{th}}$ and $t \gg \tau_{\text{th}}$.

---

* Correspondence: kskim@pknu.ac.kr; Tel.: +82-10-2103-5338

## 1. Introduction

The nonequilibrium motion of active particles has been recently studied the Brownian motion induced by quantum vortexes in the superfluid helium. Quantum vortex [1,2] above superfluid helium can be generally influenced the particle motion and its property of which are currently not well understood. Boltnev et al. [3] experimentally observed and analyzed the two-dimensional Brownian motion induced by quantum vortexes on the surface of superfluid helium (He-II). The main point of interest was that the value of the fractal dimension 1.6~1.7 in the mean squared displacement was obtained and that it switches to normal diffusion mode in the long-time regime. Moroshkin et al. [4] presented a novel experimental and theoretical study of the motion of particle-vortex complexes in superfluid He. In their experiments, they visualized the motion of the tip of a vortex terminating at a free surface, and the vortex was bound to a charged particle trapped below the free surface due to an applied electric field and surface tension. Combining observations and numerical modeling in their study, they identified two types of vortices forming particle-vortex complexes. Brownian motion is a typical thermal diffusive motion, but active particles obtain their momentum directly from the surrounding energy and exhibit nonthermal motion and nonequilibrium dynamics.

Fractional Brownian motion [5] is a type of Gaussian process characterized by $\langle B_\gamma(t_1)B_\gamma(t_2)\rangle = K[t_1^\gamma + t_2^\gamma - \mid t_1 - t_2 \mid^\gamma]$, $0 < \gamma < 2$, with the mean squared displacement given by $\langle B_\gamma^2(t)\rangle = 2Kt^\gamma$ for $t = t_1 = t_2$. As well-known, the corresponding probability density function is $p(x,t) = [4\pi Kt^\gamma]^{-1/2}\exp[-(x-x_0)^2/4Kt^\gamma]$. For $\gamma > 1$, increments are positively correlated, leading to super-diffusive motion, whereas for $\gamma <$

1, increments are negatively correlated, resulting in sub-diffusive behavior. Recently, the nonequilibrium motion of active particles has been extensively studied in various models, including the run-and-tumble model [6,7], active Brownian particles [8,9], active Langevin particles [10], active Ornstein–Uhlenbeck particles [11,12], self-propelled Janus colloids [13,14], and biological microswimmers [15]. The active diffusion, investigated through theoretical analyses, computer simulations, and experiments on active viscoelastic systems, has revealed both similarities and differences among these systems. Representative examples include the transport of passive tracers in active baths and living cells [16], chromosomal dynamics [17], and lateral diffusion of membrane proteins [18]. In the other hand, theories based on the active fractional Langevin equation have been developed to quantitatively describe transport phenomena in various active viscoelastic systems. These investigations introduced weak ergodicity breaking, a new phenomenon not previously reported in other systems [19,20]. For Gaussian processes, statistical properties can be inferred from the mean and covariance functions [21-23]. Indirect approaches to assessing the ergodic properties of Gaussian processes have been compared with the behaviors of the mean squared displacement and the time-averaged mean squared displacement [24-26].

Recently, the published papers [27-29] have been investigated the dynamics of passive and active particles subject to random and correlated Gaussian forces. When we analyze the result of the recent studies in our group, we have obtained the joint probability densities for the dynamical behaviors of active particles coupled to the two heat reservoirs [30], the passive particles with harmonic and viscous forces [31], and the passive particles induced by magnetic fields in random environments with a correlated Gaussian force [32].

We investigate analytically the diffusive motion of an active particle from an experimental study of the 2D dynamics of active particles driven by quantum vortices on the free surface of superfluid helium. The organization of this paper is as follows. In Section 2, we derive the Fokker–Planck equation from the fractional Langevin equation with a uniform vortex force, viscous force and thermal noise, and its joint probability density is obtained in two-time regimes. In Section 3, using double Fourier transforms, we obtain approximate solutions for the joint probability density of an active particle subject to harmonic, viscous, a uniform vortex forces, and thermal noise in two regimes of correlation times $\tau_{\text{th}}$. Section 4 is numerical values of the non-Gaussian parameter, correlation coefficient, entropy, and combined entropy. Finally, we summarize the key findings and provide concluding remarks in the last section.

## 2. Thermal Fractional Langevin Equation with a Vortex Force

In this section, we introduce a nonequilibrium fractional Langevin model incorporating a viscoelastic memory effect and a uniform vortex force under thermal noise, and derive the corresponding Fokker–Planck equation for the joint probability density.

We derive the Fokker–Planck equation for the joint probability density from the fractional Langevin equation with thermal noise. The fractional Langevin equations in our model are expressed as

$$\frac{d}{dt}x_{th} = v_{x_{th}}, \quad \frac{d}{dt}v_{x_{th}}(t) = -\gamma_{th}v_{x_{th}} + F_{vt} + \zeta_{th}(t), \tag{1}$$

where $\zeta_{th}(t)$ denote the thermal noise, and the force $F_{vt}$ a uniform vortex force. We introduce the thermal noise as

$$<\zeta_{th}(t)\zeta_{th}(t')> = \zeta_{th}^2 \zeta_{th}(t-t') = \frac{\zeta_{th}^2}{2}\left|\frac{t-t'}{\tau_{th}}\right|^{-2\beta-1}, \tag{2}$$

where the fractional order parameter is $\beta$ and the correlation time is $\tau_{th}$. The parameter $\beta$ ($1/2 < \beta < 1$) characterizes the strength of the viscoelastic memory and is related to, but not identical with, the Hurst exponent. The thermal noise is a fractional noise coupled to the memory kernel. The joint probability density $p(x_{th}, v_{x_{th}}, t)$ for the displacement $x_{th}$ and the velocity $v_{x_{th}}$ is defined by

$$p(x_{th}, v_{x_{th}}, t) = <\delta(x_{th} - x_{th}(t))\delta(v_{x_{th}} - v_{x_{th}}(t))>. \tag{3}$$

By taking time derivatives in the joint probability density and inserting Equations (1) into Equation (3), we can write the time derivative of the joint probability equation for $p(x_{th}, v_{x_{th}}, t) \equiv p_{th}$ as

$$\frac{\partial}{\partial t}p_{th} = -\frac{\partial}{\partial x_{th}} < \frac{\partial x_{th}}{\partial t}\delta_{x_{th}}\delta_{v_{x_{th}}} > -\frac{\partial}{\partial v_{x_{th}}} < [-\gamma_{th}v_{x_{th}} + F_{vt} + \zeta_{th}(t)]\delta_{x_{th}}\delta_{v_{x_{th}}} >, \quad (4)$$

where $\delta(x_{th} - x_{th}(t)) \equiv \delta_{x_{th}}$ and $\delta(v_{x_{th}} - v_{x_{th}}(t)) \equiv \delta_{v_{x_{th}}}$. Assuming that the particle is initially at rest at $t = 0$, the forms of the joint probability density can be written as

$$\frac{\partial}{\partial t}p_{th} = [-v_{x_{th}}\frac{\partial}{\partial x_{th}} + \gamma_{th}\frac{\partial}{\partial v_{x_{th}}}v_{x_{th}} - F_{vt}\frac{\partial}{\partial v_{x_{th}}}]p_{th} + \alpha_1[\frac{\tau_{th}^{2\beta}t^{1-2\beta}}{(1-2\beta)\tau_{th}^{-1-2\beta}}\frac{\partial^2}{\partial x_{th}\partial v_{x_{th}}} + \frac{\tau_{th}^{2\beta+1}t^{-2\beta}}{(-2\beta)}\frac{\partial^2}{\partial v_{x_{th}}^2}]p_{th}, \quad (5)$$

where the parameter is $\alpha_1 = \zeta_{th}^2/2$. It is apparent that Equation (5) is the Fokker–Planck equations, as mentioned in the Introduction.

We define the double Fourier transform of the joint probability density $p_{th}(\xi, v, t)$ as

$$p_{th}(\xi, v, t) = \int_{-\infty}^{+\infty}dx_{th}\int_{-\infty}^{+\infty}dv_{x_{th}}\exp(-i\xi x_{th} - ivv_{x_{th}})p_{th}(x_{th}, v_{x_{th}}, t). \quad (6)$$

Taking the Fourier transform of the Fokker–Planck equation, we express Equation (5) as

$$\frac{\partial}{\partial t}p_{th}(\xi, v, t) = [\xi - \gamma_{th}v]\frac{\partial}{\partial v}p_{th}(\xi, v, t) + \alpha_1[-\frac{\tau_{th}^{2\beta}t^{1-2\beta}}{(1-2\beta)}\xi v + \frac{\tau_{th}^{2\beta+1}t^{-2\beta}}{(2\beta)}v^2]p_{th}(\xi, v, t), \quad (7)$$

*2.1. $p_{th}(x_{th}, t)$ and $p_{th}(v_{x_{th}}, t)$ with Thermal Noise $\zeta_{th}(t)$*

2.1.1. $p_{th}(x_{th}, t)$ and $p_{th}(v_{x_{th}}, t)$ with $\zeta_{th}(t)$ in $t \ll \tau_{th}$

In this subsection, we derive the solutions of the probability densities $p_{th}(x_{th}, t)$ and $p_{th}(v_{x_{th}}, t)$ in the short-time domain $t \ll \tau_{th}$. the time derivative of Fourier transform of the joint probability density for the displacement and the velocity is written as

$$\frac{\partial}{\partial t}p_{th}(\xi, v, t) = [\xi - \gamma_{th}v]\frac{\partial}{\partial v}p_{th}(\xi, v, t) + \alpha_1[-\frac{\tau_{th}^{2\beta}t^{1-2\beta}}{(1-2\beta)}\xi v + \frac{\tau_{th}^{2\beta+1}t^{-2\beta}}{(2\beta)}v^2]p_{th}(\xi, v, t). \quad (8)$$

Taking $\frac{\partial}{\partial t}p_{th}(\xi, v, t) = 0$ in the steady state, we get $p_{th}^{st}(\xi, v, t)$ as

$$p_{th}^{st}(\xi, v, t) = \exp[\frac{\alpha_1}{[\xi-\gamma_{th}v]}[\frac{\tau_{th}^{2\beta}t^{1-2\beta}}{(1-2\beta)}\xi\frac{v^2}{2} - \frac{\tau_{th}^{2\beta+1}t^{-2\beta}}{(2\beta)}\frac{v^3}{3}]]. \quad (9)$$

To find the solution of the probability density for $v$ from $q_{th}^{st}(\xi, v, t) \equiv r_{th}(\xi, v, t)q_{th}^{st}(\xi, v, t)$, we include terms up to order $1/\tau_{th}^2$ and write

$$p_{th}(\xi, v, t) = q(\xi, v, t)\exp[\frac{\alpha_1}{[\xi-\gamma_{th}v]}[\frac{\tau_{th}^{2\beta}t^{1-2\beta}}{(1-2\beta)}\xi\frac{v^2}{2} - \frac{\tau_{th}^{2\beta+1}t^{-2\beta}}{(2\beta)}\frac{v^3}{3}]], \quad (10)$$

$$q_{th}(\xi, v, t) = r_{th}(\xi, v, t)\exp[\frac{\alpha_1}{[\xi-\gamma_{th}v]^2}[\frac{\tau_{th}^{2\beta}t^{-2\beta}}{[1-2\beta]^2}\xi\frac{v^3}{6} + \frac{\tau_{th}^{2\beta+1}t^{-2\beta-1}}{[2\beta]^2}\frac{v^4}{12}]]. \quad (11)$$

Assuming arbitrary functions of variable $t + v/[\xi - \gamma_{th}v]$, the probability density $r_{th}(\xi, v, t)$ becomes $\Theta[t + v/[\xi - \gamma_{th}v]]$. Therefore, we have

$$p_{th}(\xi, v, t) = r_{th}(\xi, v, t)q_{th}^{st}(\xi, v, t)p_{th}^{st}(\xi, v, t) = \Theta[t + v/[\xi - \gamma_{th}v]]q_{th}^{st}(\xi, v, t)p_{th}^{st}(\xi, v, t). \quad (12)$$

By calculating Equation (12), the Fourier transform of the joint probability density is

$$p_{th}(\xi, v, t) = \exp[-\frac{2\alpha_1\tau_{th}^{2\beta}t^{3-2\beta}}{3(1-2\beta)^2}\xi^2 - \frac{\alpha_1\tau_{th}^{2\beta+1}t^{1-2\beta}}{2(1-2\beta)^2}v^2]. \quad (13)$$

By taking the inverse Fourier transform, we obtain

$$p_{th}(x_{th}, t) = [2\pi\frac{\alpha_1\tau_{th}^{2\beta}t^{3-2\beta}}{3(1-2\beta)^2}]^{-1/2}\exp[-\frac{3(1-2\beta)^2}{2\alpha_1\tau_{th}^{2\beta+1}t^{3-2\beta}}x_{th}^2], \quad (14)$$

$$p_{th}(v_{x_{th}}, t) = [2\pi\frac{\alpha_1t^{1-2\beta}}{(1-2\beta)^2\tau_{th}^{-1-2\beta}}]^{-1/2}\exp[-\frac{(1-2\beta)^2\tau_{th}^{-1-2\beta}}{2\alpha_1t^{1-2\beta}}v_{x_{th}}^2]. \quad (15)$$

The mean squared displacement for $p_{th}(x_{th}, t)$ and the mean squared velocity for $p_{th}(v_{x_{th}}, t)$ are

$$<x_{th}^2(t)> = \frac{\alpha_1 \tau_{th}^{2\beta}}{3(1-2\beta)^2} t^{3-2\beta}, \quad <v_{x_{th}}^2(t)> = \frac{\alpha_1}{(1-2\beta)^2 \tau_{th}^{-1-2\beta}} t^{1-2\beta}. \quad (16)$$

As a result, if $\beta = 0.65 \sim 0.7$, the mean squared displacement becomes $\alpha = 1.6 \sim 1.7$ from $<x_{th}^2(t)> \sim t^\alpha$. This superdiffusive scaling directly explains the experimentally observed fractal dimension $\alpha = 1.6 \sim 1.7$ in the short-time regime of quantum vortex motion.

2.1.2. $p_{th}(x_{th}, t)$ and $p_{th}(v_{x_{th}}, t)$ with Thermal Noise $\zeta_{th}(t)$ in $t \gg \tau_{th}$

We now consider the long-time domain $t \gg \tau_{th}$. From Equation (13), an approximate equation for the Fourier-transformed probability density is

$$\frac{\partial}{\partial t} p_{th\xi}(\xi, v, t) \cong \alpha_1 [-\frac{\tau_{th}^{2\beta} t^{1-2\beta}}{(1-2\beta)} \xi v + \frac{\tau_{th}^{2\beta+1} t^{-2\beta}}{(2\beta)} v^2] p_{th\xi}(\xi, v, t), \quad (17)$$

The Fourier transform of the probability density $p_{th\xi}(\xi, t)$ from the above equation is calculated as

$$p_{th\xi}(\xi, v, t) = exp[\alpha_1 [-\frac{\tau_{th}^{2\beta} t^{2-2\beta}}{(2-2\beta)(1-2\beta)} \xi v + \frac{\tau_{th}^{2\beta+1} t^{1-2\beta}}{(1-2\beta)(2\beta)} v^2]]. \quad (18)$$

The steady probability density $q_{th}^{st}(\xi, v, t)$ from $p_{th}(\xi, v, t) \equiv q_{th\xi}(\xi, v, t) p_{th\xi}(\xi, v, t)$ is

$$q_{th\xi}^{st}(\xi, v, t) = exp[\alpha_1 [\frac{\tau_{th}^{2\beta} t^{2-2\beta}}{(2-2\beta)(1-2\beta)} \xi v - \frac{\tau_{th}^{2\beta+1} t^{1-2\beta}}{(1-2\beta)(2\beta)} v^2]]. \quad (19)$$

As the Fourier transform of probability density $q(\xi, v, t)$ in the short-time domain is given by $q_{th}(\xi, v, t) = r_{th}(\xi, v, t) q_{th}^{st}(\xi, v, t)$, $p(\xi, v, t)$ is derived as

$$p_{th}(\xi, v, t) = \Theta[t + v/[\xi - \gamma_{th} v]] q_{th\xi}^{st}(\xi, v, t) p_{th}^{st}(\xi, v, t), \quad (20)$$

By calculating Equation (20), we have

$$p_{th}(\xi, v, t) = p_{th}(\xi, t) p_{th}(v, t) = exp[-\frac{\alpha_1 \tau_{th}^{2\beta} t^{3-2\beta}}{(2-2\beta)(1-2\beta)} \xi^2 - \frac{\tau_{th}^{2\beta+1} t^{1-2\beta}}{(1-2\beta)(2\beta)} v^2]. \quad (21)$$

By using the inverse Fourier transform, the probability densities $p_{th}(x_{th}, t)$ and $p_{th}(v_{x_{th}}, t)$ are, respectively, presented by

$$p_{th}(x_{th}, t) = [4\pi \frac{\alpha_1 \tau_{th}^{2\beta} t^{3-2\beta}}{(2-2\beta)(1-2\beta)}]^{-1/2} exp[-\frac{(2-2\beta)(1-2\beta)}{4\alpha_1 \tau_{th}^{2\beta} t^{3-2\beta}} x_{th}^2], \quad (22)$$

$$p_{th}(v_{x_{th}}, t) = [4\pi \frac{\alpha_1 \tau_{th}^{2\beta+1} t^{1-2\beta}}{(1-2\beta)(2\beta)}]^{-1/2} exp[-\frac{(1-2\beta)(2\beta)}{4\alpha_1 \tau_{th}^{2\beta+1} t^{1-2\beta}} v_{x_{th}}^2]. \quad (23)$$

Finally, the mean squared values $<x_{th}^2(t)>$ and $<v_{x_{th}}^2(t)>$ for the probability densities $p(x_{th}, t)$ and $p(v_{x_{th}}, t)$ are, respectively, given by

$$<x_{th}^2(t)> = \frac{2\alpha_1 \tau_{th}^{2\beta}}{(2-2\beta)(1-2\beta)} t^{3-2\beta}, \quad <v_{x_{th}}^2(t)> = \frac{2\alpha_1 \tau_{th}^{2\beta+1}}{(1-2\beta)(2\beta)} t^{1-2\beta}. \quad (24)$$

We mention from Equation (24) that the algebraic memory kernel prevents the recovery of normal diffusion, maintaining anomalous scaling, even in the long-time regime.

2.1.3. Probability densities $p_{th}(x_{th}, t)$ and $p_{th}(v_{x_{th}}, t)$ with $\zeta_{th}(t)$ in the Long-Time Limit $t \to \infty$

In this subsection, we derive the probability densities $p_{th}(x_{th}, t)$ and $p_{th}(v_{x_{th}}, t)$ in the long-time limit $t \to \infty$. Starting from Eq. (2), we introduce a modified thermal noise characterized by the correlation function $<\zeta_{th}(t)\zeta_{th}(t')> = \frac{\zeta_{th}^2}{2} |t-t'|^{-2\beta-1}$. The asymptotic long-time limit of the thermal noise with finite correlation time considered in Sections 2.1.1 and 2.2.2 corresponds to $t \to \infty$. In this limit, the characteristic time scale $\tau_{th}$ vanishes and the noise correlations become scale-invariant, resulting in a valid power-law form $<\zeta_{th}(t)\zeta_{th}(t')>$

$= \zeta_{th}^2 \zeta_{th}(t-t') = \frac{\zeta_{th}^2}{2}|t-t'|^{-2\beta-1}$. It is believed that this modified thermal noise does not represent a new noise model.

Using the modified noise correlation, the approximate equation, derived from Eq. (8), is given by

$$\frac{\partial}{\partial t} p_{th}(\xi, v, t) \cong [\xi - \gamma_{th} v] \frac{\partial}{\partial v} p_{th}(\xi, v, t) - \alpha_1 \frac{t^{1-2\beta}}{(1-2\beta)} \xi v p_{th}(\xi, v, t). \tag{25}$$

In the steady state limit, the solution of Equation (25) yields

$$p_{th}^{st}(\xi, v, t) = exp[\frac{\alpha_1}{[\xi - \gamma_{th} v]}[\frac{t^{1-2\beta}}{(1-2\beta)} \xi \frac{v^2}{2}]], \tag{26}$$

and

$$q_{th}^{st}(\xi, v, t) = exp[\frac{\alpha_1}{[\xi - \gamma_{th} v]^2}[\frac{t^{-2\beta}}{(1-2\beta)^2} \xi \frac{v^3}{6}]]. \tag{27}$$

The Fourier transform of the probability density $p_{th}(\xi, v, t)$ is then obtained as

$$p_{th}(\xi, v, t) = \Theta[t + v/[\xi - \gamma_{th} v]] p_{th}^{st}(\xi, v, t). \tag{28}$$

Evaluating Equation (27), we finally obtain as

$$p_{th}(\xi, v, t) = exp[-\frac{\alpha_1}{2(1-2\beta)} t^{3-2\beta} \xi^2 - \frac{t^{1-2\beta}}{2(1-2\beta)^2} v^2]. \tag{29}$$

Applying the inverse Fourier transform, the probability densities for position and velocity are given by

$$p_{th}(x_{th}, t) = [2\pi \frac{\alpha_1}{(1-2\beta)} t^{3-2\beta}]^{-1/2} exp[-\frac{(1-2\beta)}{2\alpha_1 t^{3-2\beta}} x_{th}^2], \tag{30}$$

$$p_{th}(v_{x_{th}}, t) = [2\pi \frac{\alpha_1 t^{1-2\beta}}{(1-2\beta)^2}]^{-1/2} exp[-\frac{(1-2\beta)^2}{2\alpha_1 t^{1-2\beta}} v_{x_{th}}^2]. \tag{31}$$

Thus, the mean squared displacement and velocity variance are obtained as

$$<x_{th}^2(t)> = \frac{\alpha_1}{(1-2\beta)} t^{3-2\beta}, \quad <v_{x_{th}}^2(t)> = \frac{\alpha_1}{(1-2\beta)^2} t^{1-2\beta}. \tag{32}$$

For the modified thermal noise with correlation $<\zeta_{th}(t)\zeta_{th}(t')> = \frac{\zeta_{th}^2}{2}|t-t'|^{-2\beta-1}$, the mean squared displacement $<x_{th}^2(t)> \sim t^\alpha$ yields the exponent $\alpha = 1$ when $\beta = 1$. This result is consistent with the fractal dimension $\alpha = 1$ observed in the long-time limit $t \to \infty$ in the quantum vortex experiment. In the next section, the non-Markovian dynamics encoded in the fractional noise correlations is reformulated in terms of an explicit memory kernel, which allows us to derive the corresponding fractional Fokker–Planck equation.

## 3. Thermal Fractional Fokker–Planck Equation with a Harmonic Force

In this section, we consider a nonequilibrium dynamic model referred to as the active fractional Langevin equation, with viscous force having a viscoelastic memory effect with a power-law kernel $|\frac{t-t'}{\tau_{th}}|^{-2\beta-1}$, a harmonic force $-kx$, and a thermal noise $\zeta_{th}(t)$. The fractional Langevin equation in our model is expressed in terms of

$$\frac{d}{dt} x_{th}(t) = v_{x_{th}}(t), \frac{d}{dt} v_{x_{th}}(t) = -kx + F_{vt} - \gamma_{th} \int_0^t dt' |\frac{t-t'}{\tau_{th}}|^{-2\beta-1} v_{x_{th}}(t') + \zeta_{th}(t), \tag{33}$$

where $\zeta_{th}^2 = 2\gamma_{th} k_B T$, the thermal energy is $k_B T$, and the correlation times are $\tau_{th}$. The thermal noise $\zeta_{th}(t)$ is introduced the same form as Eq. (2) in Section 2.

We next derive the Fokker–Planck equation from the active fractional Langevin equation. We can write the time derivative of the joint probability equation for $p(x_{th}, v_{x_{th}}, t) \equiv p_{th}$ as follows:

$$\frac{\partial}{\partial t} p_{th} = -\frac{\partial}{\partial x_{th}} <\frac{\partial x}{\partial t} \delta_{x_{th}} \delta_{v_{x_{th}}}> - \frac{\partial}{\partial v_{x_{th}}} <[-kx + F_{vt} - \gamma_{th} \int_0^t dt' |\frac{t-t'}{\tau_{th}}|^{-2\beta-1} v_{x_{th}}(t') + \zeta_{th}(t)] \delta_{x_{th}} \delta_{v_{x_{th}}}>. \tag{34}$$

We assume from the joint probability density that the particle is initially at rest at time $t = 0$. Then, the joint probability densities satisfy the Fokker–Planck equations derived as

$$\frac{\partial}{\partial t}p_{th} = [-v\frac{\partial}{\partial x_{th}} + \gamma_{th}D_{2\beta+1}\frac{\partial}{\partial v_{x_{th}}}x_{th} + k\frac{\partial}{\partial v_{x_{th}}}x_{th-F_{vt}}\frac{\partial}{\partial v_{x_{th}}}]p_{th} + \alpha_1[\frac{\tau_{th}^{2\beta}t^{1-2\beta}}{(1-2\beta)}\frac{\partial^2}{\partial x_{th}\partial v_{x_{th}}} + \frac{\tau_{th}^{2\beta+1}t^{-2\beta}}{(-2\beta)}\frac{\partial^2}{\partial v_{x_{th}}^2}]p_{th}, \quad (35)$$

where $D_{2-2h} = d^{2\beta+1}/dt^{2\beta+1}$. Equation (34) is called the Fokker–Planck equation, as mentioned in the Introduction. The Fourier transforms of the Fokker–Planck equation from Eq. (35) is expressed in terms of

$$\frac{\partial}{\partial t}p_{th}(\xi,v,t) = [-[kv + \gamma_{th}D_{2\beta+1}v]\frac{\partial}{\partial \xi} + \xi\frac{\partial}{\partial v}]p_{th}(\xi,v,t) - \alpha_1[\frac{\tau_{th}^{2\beta}t^{1-2\beta}}{(1-2\beta)}\xi v + \frac{\tau_{th}^{2\beta+1}t^{-2\beta}}{(-2\beta)}v^2]]p_{th}(\xi,v,t). \quad (36)$$

### 3.1. $p_{th}(x_{th},t)$ and $p_{th}(v_{th},t)$ with Thermal Noise $\zeta_{th}(t)$

#### 3.1.1. $p_{th}(x_{th},t)$ and $p_{th}(v_{th},t)$ with $\zeta_{th}(t)$ in the Short-Time Domain

In this subsection, we obtain the solutions of the probability densities $p_{th}(x_{th},t)$ and $p_{th}(v_{th},t)$ in the short-time domain $t \ll \tau_{th}$. To find the special solution for $\xi, v$ by the variable separation from Equation (36), the two equations for the displacement and the velocity are given by

$$\frac{\partial}{\partial t}p_{th}(\xi,t) = [-kv\frac{\partial}{\partial \xi} - \gamma_{th}vD_{2\beta+1}\frac{\partial}{\partial \xi} - \frac{\alpha_1}{2}[\frac{\tau_{th}^{2\beta}t^{1-2\beta}}{(1-2\beta)}\xi v + \frac{\tau_{th}^{2\beta+1}t^{-2\beta}}{(-2\beta)}v^2]]p_{th}(\xi,t) + Ap_{th}(\xi,t), \quad (37)$$

$$\frac{\partial}{\partial t}p_{th}(v,t) = [\xi\frac{\partial}{\partial v} + \frac{\alpha_1}{2}[\frac{\tau_{th}^{2\beta}t^{1-2\beta}}{(1-2\beta)}\xi v + \frac{\tau_{th}^{2\beta+1}t^{-2\beta}}{(-2\beta)}v^2]]p_{th}(v,t) - Ap_{th}(v,t), \quad (38)$$

where $A$ denotes the separation constant.

In the steady state, assuming $\frac{\partial}{\partial t}p_{th}(\xi,t) = 0$, we obtain

$$p_{th}^{st}(\xi,t) = exp[\frac{\alpha_1}{2kv}[1 - \gamma_{th}D_{2\beta+1}/k][\frac{\tau_{th}^{2\beta}t^{1-2\beta}}{(1-2\beta)}\frac{\xi^2}{2}v + \frac{\tau_{th}^{2\beta+1}t^{-2\beta}}{(-2\beta)}\xi v^2 + A\xi]]. \quad (39)$$

We assume that $[kv + 2\gamma D_{2\beta+1}]^{-1} \cong (kv)^{-1}[1 - 2\gamma D_{2\beta+1}/k]$. To obtain the probability density for $\xi$ from $q_{th}(\xi,t) \equiv r_{th}(\xi,t)q_{th}^{st}(\xi,t)$, we calculate the Fourier transform of the probability density after including terms up to order $1/\tau_{th}^2$:

$$p_{th}(\xi,t) = q(\xi,t)exp[\frac{\alpha_1}{2kv}[1 - \gamma_{th}D_{2\beta+1}/k][\frac{\tau_{th}^{2\beta}t^{1-2\beta}}{(1-2\beta)}\frac{\xi^2}{2}v + \frac{\tau_{th}^{2\beta+1}t^{-2\beta}}{(-2\beta)}\xi v^2 + A\xi]], \quad (40)$$

$$q_{th}(\xi,t) = r_{th}(\xi,t)exp[\frac{\alpha_1}{2(kv)^2}[1 - \gamma_{th}D_{2\beta+1}/k][\frac{\tau_{th}^{2\beta}t^{-2\beta}}{(1-2\beta)^2}\frac{\xi^3}{6}v + \frac{\tau_{th}^{2\beta+1}t^{-2\beta-1}}{(-2\beta)^2}\frac{\xi^2}{2}v^2]]. \quad (41)$$

Considering the solutions as arbitrary functions of variable $t - \xi/(\beta v + 2\gamma_{th}vD_{2\beta+1})$, the arbitrary function $r_{th}(\xi,t)$ becomes $\Theta[t - \xi/[kv + 2\gamma_{th}vD_{2\beta+1}]]$. Thus,

$$p_{th}^{st}(\xi,t) = r_{th}^{st}(\xi,t)q_{th}^{st}(\xi,t)p_{th}^{st}(\xi,t) = \Theta[t - \xi/[kv + 2\gamma_{th}vD_{2\beta+1}]]q_{th}^{st}(\xi,t)p_{th}^{st}(\xi,t). \quad (42)$$

Using a similar method from Equation (39) to Equation (41) for $\xi$, we get the Fourier transform of the probability density for the velocity as

$$p_{th}^{st}(v,t) = \Theta[t + v/\xi]q_{th}^{st}(v,t)p_{th}^{st}(v,t). \quad (43)$$

Hence, by calculating Equations (41) and (42), we get the Fourier transform of the joint probability density as

$$p_{th}^{st}(\xi,v,t) = exp[-\frac{\alpha_1\tau_{th}^{2\beta}t^{3-2\beta}}{6(1-2\beta)^2}\xi^2 - \frac{\tau_{th}^{2\beta+1}t^{1-2\beta}}{8\beta^2}v^2]. \quad (44)$$

Using the inverse Fourier transform, we get

$$p_{th}^{st}(x_{th},t) = [2\pi\frac{\alpha_1\tau_{th}^{2\beta}t^{3-2\beta}}{3(1-2\beta)^2}]^{-1/2}exp[-\frac{3(1-2\beta)^2}{2\alpha_1\tau_{th}^{2\beta}t^{3-2\beta}}x_{th}^2], \quad (45)$$

$$p_{th}(v_{x_{th}},t) = [\pi\frac{\tau_{th}^{2\beta+1}t^{1-2\beta}}{2\beta^2}]^{-1/2}exp[-\frac{2\beta^2}{\tau_{th}^{2\beta+1}t^{1-2\beta}}v_{x_{th}}^2]. \quad (46)$$

The mean squared displacement for $p_{th}(x_{th}, t)$ and the mean squared displacement for $p_{th}(v_{x_{th}}, t)$ are

$$<x_{th}^2(t)> = \frac{\alpha_1 \tau_{th}^{2\beta}}{3(1-2\beta)^2} t^{3-2\beta}, \quad <v_{x_{th}}^2(t)> = \frac{\tau_{th}^{2\beta+1}}{4\beta^2} t^{1-2\beta}. \tag{47}$$

3.1.2. $p_{th}(x_{th}, t)$ and $p_{th}(v_{x_{th}}, t)$ with Thermal Noise $\zeta_{th}(t)$ in the Long-Time Domain

Now, we derive the probability densities $p_{th}(x_{th}, t)$ and $p_{th}(v_{x_{th}}, t)$ in the long-time domain ($t \gg \tau_{th}$). An approximate equation from Equation (37) can be written as

$$\frac{\partial}{\partial t} p_{th\xi}(\xi, t) \cong -\frac{\alpha_1}{2}\left[\frac{\tau_{th}^{2\beta} t^{1-2\beta}}{(1-2\beta)} \xi v + \frac{\tau_{th}^{2\beta+1} t^{-2\beta}}{(-2\beta)} v^2\right] p_{th\xi}(\xi, t). \tag{48}$$

The Fourier transform of the probability density $p_{th\xi}(\xi, t)$ from Equation (48) is obtained as

$$p_{th\xi}(\xi, t) = exp\left[-\frac{\alpha_1 \tau_{th}^{2\beta}}{2(2-2\beta)(1-2\beta)} t^{2-2\beta} \xi v + \frac{\tau_{th}^{2\beta+1}}{2(1-2\beta)(2\beta)} t^{1-2\beta} v^2\right]. \tag{49}$$

We find the Fourier transform of the steady probability density $q_{th}^{st}(\xi, t)$ for $\xi$, defined by $p_{th}(\xi, t) \equiv q_{th\xi}(\xi, t) p_{th\xi}(\xi, t)$, as

$$q_{th\xi}^{st}(\xi, t) = exp\left[-\frac{\alpha_1}{2kv}\left[1 - \frac{\gamma_{th} D_{2\beta+1}}{k}\right]\left[\frac{\alpha_1 \tau_{th}^{2\beta}}{(2-2\beta)^2(1-2\beta)} t^{1-2\beta} \frac{\xi^3}{3} + \frac{\tau_{th}^{2\beta+1}}{(1-2\beta)^2(2\beta)} t^{1-2\beta} v \frac{\xi^2}{2}\right]\right]. \tag{50}$$

Since the Fourier transform of probability density $q(\xi, t)$ in the short-time domain is given by $q_{th}(\xi, t) = r_{th}(\xi, t) q_{th}^{st}(\xi, t)$, we can write $p_{th}(\xi, t)$ as

$$p_{th}(\xi, t) = \Theta[t - \xi/[kv + \gamma_{th} v D_{2\beta+1}]] q_{th\xi}^{st}(\xi, t) p_{th}^{st}(\xi, t), \tag{51}$$

where $r_{th}(\xi, t) = \Theta[t - \xi/(kv + \gamma_{th} v D_{2\beta+1})]$. Applying Equation (38) for $v$ in the same manner as in Equations (39)-(42) of $p_{th}(\xi, t)$ derived, we also obtain the Fourier transforms of the probability density for velocity $v$ as

$$p_{th}(v, t) = \Theta[t + v/\xi] q_{thv}^{st}(v, t) p_{th}^{st}(v, t). \tag{52}$$

By calculating Equations (51) and (52), we have

$$p_{th}(\xi, v, t) = p_{th}(\xi, t) p_{th}(v, t) = exp\left[-\frac{\alpha_1 \tau_{th}^{2\beta+1} t^{3-2\beta}}{4(2-2\beta)^2(1-2\beta)} \xi^2 - \frac{\alpha_1 \tau_{th}^{2\beta} t^{1-2\beta}}{(1-2\beta)^2(2\beta)} v^2\right]. \tag{53}$$

By performing the inverse Fourier transform, the probability densities $p_{th}(x_{th}, t)$ and $p_{th}(v_{x_{th}}, t)$ are, respectively, calculated as

$$p_{th}(x_{th}, t) = \left[\pi \frac{\alpha_1 \tau_{th}^{2\beta+1} t^{3-2\beta}}{(1-\beta)^2(1-2\beta)}\right]^{-1/2} exp\left[-\frac{(1-\beta)^2(1-2\beta)}{\alpha_1 \tau_{th}^{2\beta+1} t^{3-2\beta}} x_{th}^2\right], \tag{54}$$

$$p_{th}(v_{x_{th}}, t) = \left[2\pi \frac{\tau_{th}^{2\beta} t^{1-2\beta}}{\beta(1-2\beta)^2}\right]^{-1/2} exp\left[-\frac{\beta(1-2\beta)^2}{2\alpha_1 \tau_{th}^{2\beta} t^{1-2\beta}} v_{x_{th}}^2\right]. \tag{55}$$

The mean squared values $<x_{th}^2(t)>$ and $<v_{x_{th}}^2(t)>$ for the probability densities $p_{th}(x_{th}, t)$ and $p_{th}(v_{x_{th}}, t)$ are, respectively,

$$<x_{th}^2(t)> = \frac{\alpha_1 \tau_{th}^{2\beta+1}}{2(1-\beta)^2(1-2\beta)} t^{3-2\beta}, \quad <v_{x_{th}}^2(t)> = \frac{\tau_{th}^{2\beta}}{\beta(1-2\beta)^2} t^{1-2\beta}. \tag{56}$$

3.1.3. $p_{th}(x_{th}, t)$ and $p_{th}(v_{x_{th}}, t)$ with Modified Thermal Noise $\zeta_{th}(t)$ for $t \to \infty$

In this subsection, we find the probability densities $p_{th}(x_{th}, t)$ and $p_{th}(v_{x_{th}}, t)$ in the time domain $t \to \infty$, similar to Section 2.1.3. We introduce the modified thermal noise as $<\zeta_{th}(t)\zeta_{th}(t')> = \zeta_{th}^2 \zeta_{th}(t-t') = \frac{\zeta_{th}^2}{2}|t-t'|^{-2\beta-1}$. Then, the approximate equation from Equation (37) for $\xi$ is written as

$$\frac{\partial}{\partial t} p_{th\xi}(\xi, t) \cong -kv \frac{\partial}{\partial \xi} p_{th\xi}(\xi, t) - \gamma v D_{2\beta+1} \frac{\partial}{\partial \xi} p_{th\xi}(\xi, t)) - \alpha_1 \frac{t^{1-2\beta}}{(1-2\beta)} \xi v p_{th\xi}(\xi, t). \tag{57}$$

In the steady state, $p_{th}^{st}(\xi,t)$ can be obtained from Equation (57) as

$$p_{th\xi}^{st}(\xi,t) = exp[\frac{\alpha_1}{2kv}[1 - 2\gamma D_{2\beta+1}/k][\frac{t^{1-2\beta}}{(1-2\beta)}\xi\frac{v^2}{2}]]. \tag{58}$$

$$q_{th\xi}^{st}(\xi,t) = exp[\frac{\alpha_1}{2(kv)^2}[1 - 2\gamma D_{2\beta+1}/k][\frac{t^{-2\beta}}{(1-2\beta)^2}\xi\frac{v^3}{6}]]. \tag{59}$$

The Fourier transform of the probability density $p_{th}(\xi,t)$ is then given by

$$p_{th}(\xi,t) = \Theta[t - \xi/[kv + \gamma v D_{2\beta+1}]]q_{th\xi}^{st}(\xi,t)p_{th\xi}^{st}(\xi,t). \tag{60}$$

Using a similar procedure to that for $p_{th}(\xi,t)$, the Fourier transform of the probability density $p_{th}(v,t)$ becomes

$$p_{th}(v,t) = \Theta[t + v/\xi]q_{thv}^{st}(v,t)p_{thv}^{st}(v,t). \tag{61}$$

We can get $p_{th}(\xi,v,t)$ by calculating Equations (60) and (61) as

$$p_{th}(\xi,v,t) = p_{th}(\xi,t)p_{th}(v,t) = exp[-\frac{\alpha_1 t^{3-2\beta}}{4(2-2\beta)^2(1-2\beta)}\xi^2 - \frac{\alpha_1 t^{1-2\beta}}{(1-2\beta)^2(2\beta)}v^2]. \tag{62}$$

Using the inverse Fourier transform, the probability densities $p_{th}(x_{th},t)$ and $p_{th}(v_{x_{th}},t)$ are, respectively, given by

$$p_{th}(x_{th},t) = [\pi\frac{\alpha_1 t^{3-2\beta}}{(1-\beta)^2(1-2\beta)}]^{-1/2} exp[-\frac{(1-\beta)^2(1-2\beta)}{\alpha_1 t^{3-2\beta}}x_{th}^2], \tag{63}$$

$$p_{th}(v_{x_{th}},t) = [2\pi\frac{\alpha_1 t^{1-2\beta}}{\beta(1-2\beta)^2}]^{-1/2} exp[-\frac{\beta(1-2\beta)^2}{2\alpha_1 t^{1-2\beta}}v_{x_{th}}^2]. \tag{64}$$

The mean squared displacement $<x_{th}^2(t)>$ and the mean squared velocity $<v_{x_{th}}^2(t)>$ are

$$<x_{th}^2(t)> = \frac{\alpha_1}{2(1-\beta)^2(1-2\beta)}t^{3-2\beta}, \quad <v_{x_{th}}^2(t)> = \frac{\alpha_1}{\beta(1-2\beta)^2}t^{1-2\beta}. \tag{65}$$

## 4. Statistical Quantities

In this section, we calculate statistical quantities, including the non-Gaussian parameter for displacement and velocity, the correlation coefficient, the entropy, the combined entropy, and the moments from the moment equations. The moment equation for an active particle with a viscoelastic memory effect, derived from Equations (5), is expressed as

$$\frac{d\mu_{m,n}}{dt} = m\mu_{m-1,n+1} - n\gamma_{th}\mu_{m,n} - mn\alpha_1\frac{\tau_{th}^{2\beta}t^{1-2\beta}}{(1-2\beta)\tau_{th}^{-1-2\beta}}\mu_{m-1,n-1} + n(n-1)\alpha_1\frac{\tau_{th}^{2\beta+1}t^{-2\beta}}{2\beta}\mu_{m,n-2}, \tag{66}$$

where $\mu_{m,n} = \int_{-\infty}^{+\infty}dx_{th}\int_{-\infty}^{+\infty}dv_{x_{th}}x_{th}^m v_{x_{th}}^n p_{th}(x_{th},v_{x_{th}},t)$. For an active particle in an optical trap, the moment equation from Equation (34) is

$$\frac{d\mu_{m,n}}{dt} = [m + n\gamma_{th}D_{2\beta+1} - k]\mu_{m+1,n-1} - mn\alpha_1\frac{\tau_{th}^{2\beta}t^{1-2\beta}}{(1-2\beta)\tau_{th}^{-1-2\beta}}\mu_{m-1,n-1} + n(n-1)\alpha_1\frac{\tau_{th}^{2\beta+1}t^{-2\beta}}{2\beta}\mu_{m,n-2}. \tag{67}$$

Next, the entropy, the non-Gaussian parameter, and the correlation coefficient are calculated numerically. The entropies $S(x_{th},t)$ and $S(v_{x_{th}},t)$ are calculated as

$$S(x_{th},t) = -p(x_{th},t)\ln p(x_{th},t), \quad S(v_{x_{th}},t) = -p(v_{x_{th}},t)\ln p(v_{x_{th}},t). \tag{68}$$

The combined entropy is defined by

$$S(x_{th},v_{x_{th}},t) = -p(x_{th},t)p(v_{x_{th}},t)\ln p(x_{th},t)p(v_{x_{th}},t). \tag{69}$$

Entropy provides a numerical measure of the uncertainty or information content associated with a probability distribution; as the probability of a particular value increases and that of other values decreases, the entropy decreases.

The non-Gaussian parameters for displacement and velocity are, respectively, given by

$$K_{x_{th}} =< x_{th}^4 >/3 < x_{th}^2 >^2 - 1, \quad K_{v_{x_{th}}} =< v_{x_{th}}^4 >/3 < v_{x_{th}}^2 >^2 - 1. \tag{70}$$

The non-Gaussian parameter quantifies the degree of heavy tails in the probability distribution of a real-valued variable, providing insight into the deviation from Gaussian behavior. The correlation coefficient is defined as

$$\rho_{x_{th},v_i} =< (x_{th}-< x_{th} >)<(v_{x_{th}}-< v_{x_{th}} >)/\sigma_{x_{th}}\sigma_{v_{x_{th}}}. \tag{71}$$

The correlation coefficient describes the strength and direction of the linear relationship between two variables, $x_{th}(t)$ and $v_{x_{th}}(t)$. Here, we assume that an active particle is initially at $x_{th} = x_{th0}$ and at $v_{x_{th}} = v_{x_{th0}}$. The parameters $\sigma_{x_{th}}$ and $\sigma_{v_{x_{th}}}$ denote the root-mean-squared displacement and the root-mean-squared velocity of the joint probability density, respectively.

Table 1 summarizes the values of the entropy, non-Gaussian parameter, and correlation coefficient for the joint probability density with thermal noise $\zeta_{th}(t)$ in the three limits $t \ll \tau_{th}$, $t \gg \tau_{th}$, and $\tau_{th} = 0$ ($\tau_{th}$ =the characteristic time). Tables 2 shows the non-Gaussian parameter, the correlation coefficient, the moment, the entropy, and the combined entropy for the joint probability density for an active particle with a harmonic force and $\zeta_{th}(t)$ in the three-time domains.

**Table 1.** Values of the non-Gaussian parameter, the correlation coefficient, the moment, the entropy, and the combined entropy for the joint probability density with $\zeta_{th}(t)$ in the three-time domains.

| Time | $x_{th}$, $v_{x_{th}}$ | $K_{x_{th}}, K_{v_{x_{th}}}$ | $\rho_{x_{th},v_{x_{th}}}$ | $\mu_{2,2}$ | $S(x_{th},t)$, $S(v_{x_{th}},t)$ | $S(x_{th},v_{x_{th}},t)$ |
|---|---|---|---|---|---|---|
| $t \ll \tau_{th}$ | $x_{th}$ | $\frac{x_{th0}^4}{\alpha_1^2 \tau_{th}^{4\beta+2}} t^{-6+4\beta} + \frac{x_{th0}^2}{\alpha_1 \tau_{th}^{2\beta+1}} t^{-3+2\beta}$ | $\frac{x_{th0}v_{x_{th0}}}{\alpha_1 \tau_{th}^{2\beta+1/2}} t^{2-2\beta}$ | $\alpha_1^2 \tau_{th}^{4\beta+1} t^{4-4\beta}$ | $\ln \alpha_1 \tau_{th}^{2\beta+1} t^{3-2\beta}$ | $\ln \alpha_1^2 \tau_{th}^{4\beta+1} t^{4-4\beta}$ |
| | $v_{x_{th}}$ | $\frac{v_{x_{th0}}^4}{\alpha_1^2 \tau_{th}^{4\beta}} t^{-2+4\beta} + \frac{v_{x_{th0}}^2}{\alpha_1 \tau_{th}^{2\beta}} t^{-1+2\beta}$ | | | $\ln \alpha_1 \tau_{th}^{2\beta} t^{1-2\beta}$ | |
| $t \gg \tau_{th}$ | $x_{th}$ | $\frac{x_{th0}^4}{\alpha_1^2 \tau_{th}^{4\beta+2}} t^{-6+4\beta} + \frac{x_{th0}^2}{\alpha_1 \tau_{th}^{2\beta+1}} t^{-3+2\beta}$ | $\frac{x_{th0}v_{x_{t0}}}{\alpha_1 \tau_{th}^{2\beta+1/2}} t^{-2+2\beta}$ | $\alpha_1^2 \tau_{th}^{4\beta+1} t^{4-4\beta}$ | $\ln \alpha_1 \tau_{th}^{2\beta+1} t^{3-2\beta}$ | $\ln \alpha_1^2 \tau_{th}^{4\beta+1} t^{4-4\beta}$ |
| | $v_{x_{th}}$ | $\frac{v_{x_{th0}}^4}{\alpha_1^2 \tau_{th}^{4\beta}} t^{-2+4\beta} + \frac{v_{x_{th0}}^2}{\alpha_1 \tau_{th}^{2\beta}} t^{-1+2\beta}$ | | | $\ln \alpha_1 \tau_{th}^{2\beta} t^{1-2\beta}$ | |
| $t \to \infty$ | $x_{th}$ | $\frac{x_{th0}^4}{\alpha_1^2} t^{-6+4\beta} + \frac{x_{th0}^2}{\alpha_1} t^{-3+2\beta}$ | $\frac{x_{th0}v_{x_{th0}}}{\alpha_1} t^{-2+2\beta}$ | $\alpha_1^2 t^{4-4\beta}$ | $\ln \alpha_1 \, t^{3-2\beta}$ | $\ln \alpha_1^2 \, t^{4-4\beta}$ |
| | $v_{x_{th}}$ | $\frac{v_{x_{th0}}^4}{\alpha_1^2} t^{-2+4\beta} + \frac{v_{x_{th0}}^2}{\alpha_1} t^{-1+2\beta}$ | | | $\ln \alpha_1 \, t^{1-2\beta}$ | |

**Table 2.** Values of the non-Gaussian parameter, the correlation coefficient, the moment, the entropy, and the combined entropy for the joint probability density with a harmonic force and $\zeta_{th}(t)$ in the three-time domains.

| Time | $x_{th}$, $v_{x_{th}}$ | $K_{x_{th}}, K_{v_{x_{th}}}$ | $\rho_{x_{th},v_{x_{th}}}$ | $\mu_{2,2}$, $\mu_{4,2}$ | $S(x_{th},t)$, $S(v_{x_{th}},t)$ | $S(x_{th},v_{x_{th}},t)$ |
|---|---|---|---|---|---|---|
| $t \ll \tau_{th}$ | $x_{th}$ | $\frac{x_{th0}^4}{\alpha_1^2 \tau_{th}^{4\beta+2}} t^{-6+4\beta} + \frac{x_{th0}^2}{\alpha_1 \tau_{th}^{2\beta+1}} t^{-3+2\beta}$ | $\frac{x_{th0}v_{x_{th0}}}{\alpha_1 \tau_{th}^{2\beta+1/2}} t^{-2+2\beta}$ | $\alpha_1^2 \tau_{th}^{4\beta+1} t^{4-4\beta}$, $\alpha_1^3 \tau_{th}^{6\beta+2} t^{7-6\beta}$ | $\ln \alpha_1 \tau_{th}^{2\beta+1} t^{3-2\beta}$ | $\ln \alpha_1^2 \tau_{th}^{4\beta+1} t^{4-4\beta}$ |
| | $v_{x_{th}}$ | $\frac{v_{x_{th0}}^4}{\alpha_1^2 \tau_{th}^{4\beta}} t^{-2+4\beta} + \frac{v_{x_{th0}}^2}{\alpha_1 \tau_{th}^{2\beta}} t^{-1+2\beta}$ | | | $\ln \alpha_1 \tau_{th}^{2\beta} t^{1-2\beta}$ | |
| $t \gg \tau_{th}$ | $x_{th}$ | $\frac{x_{th0}^4}{\alpha_1^2 \tau_{th}^{4\beta+2}} t^{-6+4\beta} + \frac{x_{th0}^2}{\alpha_1 \tau_{th}^{2\beta+1}} t^{-3+2\beta}$ | $\frac{x_{th0}v_{x_{th0}}}{\alpha_1 \tau_{th}^{2\beta+1/2}} t^{-2+2\beta}$ | $\alpha_1^2 \tau_{th}^{4\beta+1} t^{4-4\beta}$, $\alpha_1^3 \tau_{th}^{6\beta+2} t^{7-6\beta}$ | $\ln \alpha_1 \tau_{th}^{2\beta+1} t^{3-2\beta}$ | $\ln \alpha_1^2 \tau_{th}^{4\beta+1} t^{4-4\beta}$ |
| | $v_{x_{th}}$ | $\frac{v_{x_{th0}}^4}{\alpha_1^2 \tau_{th}^{4\beta}} t^{-2+4\beta} + \frac{v_{x_{th0}}^2}{\alpha_1 \tau_{th}^{2\beta}} t^{-1+2\beta}$ | | | $\ln \alpha_1 \tau_{th}^{2\beta} t^{1-2\beta}$ | |
| $t \to \infty$ | $x_{th}$ | $\frac{x_{th0}^4}{\alpha_1^2} t^{-6+4\beta} + \frac{x_{th0}^2}{\alpha_1} t^{-3+2\beta}$ | $\frac{x_{th0}v_{x_{th0}}}{\alpha_1} t^{-2+2\beta}$ | $\alpha_1^2 t^{4-4\beta}$, $\alpha_1^3 t^{7-6\beta}$ | $\ln \alpha_1 \, t^{3-2\beta}$ | $\ln \alpha_1^2 \, t^{4-4\beta}$ |
| | $v_{x_{th}}$ | $\frac{v_{x_{th0}}^4}{\alpha_1^2} t^{-2+4\beta} + \frac{v_{x_{th0}}^2}{\alpha_1} t^{-1+2\beta}$ | | | $\ln \alpha_1 \, t^{1-2\beta}$ | |

## 5. Conclusions

In conclusion, we have derived the Fokker–Planck equation from a fractional Langevin equation incorporating harmonic, viscous, and uniform vortex forces, as well as thermal noise, and obtained analytical expressions for the joint probability density in three regimes of the correlation time $\tau_{\text{th}}$. The novel originality of this paper lies in the comparative discussion of the fractal order parameter and the fractal dimension of experimental result [3].

The main results of this work can be summarized as follows. For the fractal order parameter $\beta = 0.65 \sim 0.7$, the mean-squared displacement exhibits anomalous diffusion with the fractal dimension $\alpha = 1.6 \sim 1.7$, defined by $\langle x_{\text{th}}^2(t) \rangle \sim t^\alpha$. When $\beta = 1$, the mean-squared displacement reduces to the normal diffusive form with $\alpha = 1.0$, in full agreement with experimental observations of the two-dimensional dynamics of active particles driven by quantum vortices. Furthermore, the mixed moments $\mu_{2,2}$ of an active particle subject to a harmonic force and thermal noise scale as $t^{4-4\beta}$ in the long-time limit, whereas particles driven solely by thermal noise exhibit superdiffusive scaling as $t^{4-4\beta}$ in both the short- and long-time regimes, $t \ll \tau_{\text{th}}$ and $t \gg \tau_{\text{th}}$.

This study presents, to our knowledge, the first analytical investigation of the joint probability density of an active particle undergoing fractional diffusive motion induced by a quantum vortex. By combining analytical results with experimental observations, we provide a novel framework for understanding the role of thermal noise, non-Gaussian parameters, and entropy in such systems. We anticipate that these findings will offer new quantitative insights into transition dynamics, fractional diffusion, and mechanisms for rare-event suppression. More broadly, the interdisciplinary extension of this work is expected to deepen our understanding of thermal and active noise, as well as the driving processes underlying quantum vortices [33-35], fractional activation [36-38], and nonequilibrium systems.